\documentclass[12pt]{revtex4}
\usepackage{dcolumn}
\usepackage{graphicx}
\usepackage{amsmath}
\usepackage{amsfonts}
\usepackage{amssymb}
\usepackage{psfrag}
\usepackage{wrapfig}
\usepackage{subfigure}
\usepackage{makeidx}
\usepackage{bm}
\usepackage{epsf}
\usepackage{hyperref}

\newcommand{\bea}{\begin{eqnarray}}
\newcommand{\eea}{\end{eqnarray}}
\newcommand{\bean}{\begin{eqnarray*}}
\newcommand{\eean}{\end{eqnarray*}}
\newcommand{\nn}{\nonumber \\}

\def\W #1{\widetilde{#1}}

\def\eref#1{(\ref{#1})}

\def\la{\lambda}

\def\Label#1{\label{#1}%
  \smash{\hbox to0pt{\raise1ex\hbox{\tiny[#1]}\hss}}}

\begin{document}

\title{Analytic expressions of amplitudes by the cross-ratio identity method}

\author{Kang Zhou$^a$}

\address{$^{a}$School of Mathematical Sciences, Zhejiang
University, Hangzhou, 310027, China}

\begin{abstract}

In order to obtain the analytic expression of an amplitude from
a generic CHY-integrand, a new algorithm based on the so-called
cross-ratio identities has been proposed recently.
In this paper, we apply this new approach to
a variety of theories including: non-linear sigma model, special Galileon
theory, pure Yang-Mills theory, pure gravity,
Born-Infeld theory, Dirac-Born-Infeld theory and
its extension,
Yang-Mills-scalar theory, Einstein-Maxwell theory
as well as Einstein-Yang-Mills theory. CHY-integrands
of these theories which contain higher-order poles
can be calculated conveniently by using the cross-ratio
identity method, and all results above have been verified numerically.

\end{abstract}


\maketitle

\section{Introduction}

In the past a few years, an elegant new formulation of the tree-level S-matrix
in arbitrary dimensions for a wide range of field theories has been presented by Cachazo, He and Yuan (CHY) \cite{Cachazo:2013gna,Cachazo:2013hca,Cachazo:2013iea,Cachazo:2014nsa,Cachazo:2014xea}.
This formulation describes the scattering amplitude for $n$ massless particles as a multidimensional contour
integral over the moduli space of punctured Riemann spheres ${\cal M}_{0,n}$.
It can be unified into a concise expression
\bea \mathcal{A}_{n}&=&\int\, \Big( {\prod_{i=1}^nd z_i\over
\mbox{vol}~SL(2,\mathbb{C})}\Big)\,\Big({\prod}'\delta(\mathcal{E}_i)\Big)\,{\cal I}_n(k,\epsilon,
z)\nn
&=&\int\,
\Big(z_{rs}z_{st}z_{tr}\prod_{i\in\{1,2,\ldots,n\}\setminus\{r,s,t\}}
dz_i\Big)\,\Big({z_{ab}z_{bc}z_{ca}}\prod_{i\in\{1,2,\ldots,n\}\setminus\{a,b,c\}}\delta(\mathcal{E}_i)\Big)\,{\cal I}_n(k,\epsilon,
z)\,,~~~\label{CHYfoumula}\eea
where $z_i$ is the puncture location in $\mathbb{CP}^1$ for the $i$-th particle,
and $z_{ij}$ is defined as $z_{ij}\equiv z_i-z_j$. The second line in \eref{CHYfoumula}
is obtained by fixing the gauge redundancy of
the M\"obius $SL(2,\mathbb{C})$ group. The $\delta$-functions impose
the scattering equations
\bea
\mathcal{E}_i\equiv\sum_{j\in\{1,2,\ldots,n\}\setminus\{i\}}{s_{ij}\over
z_{ij}}=0\,,~~~\label{seqn}\eea
where $s_{ij}\equiv(k_i+k_j)^2=2k_i\cdot k_j$ are the ordinary Mandelstam variables
(in general, we use $s_{ij\cdot\cdot\cdot k}\equiv(k_i+k_j+\cdot\cdot\cdot k_k)^2$ conventionally).
These equations fully localize the integration to a sum over $(n-3)!$ solutions,
and no actual integration is required for calculating the $n$-point amplitude.
The M\"obius invariant integrand ${\cal I}_n(k,\epsilon,
z)$ is a rational function of complex variables $z_i$'s, external momenta $k_i$'s
and polarization vectors $\epsilon_i$'s.
${\cal I}_n(k,\epsilon,
z)$ depends on the theory under consideration and carries all the information
about wave functions of external particles.

Although conceptually simple and elegant, when applied to practical evaluations,
the essential step of finding the full set of analytic solutions becomes a major obstacle,
due to the Abel-Ruffini theorem that there is no algebraic solution to the general
polynomial equations of degree five or higher with arbitrary coefficients.
Moreover, after summing over all those solutions, one often ends up with
a remarkably simple result. It is natural to wonder if there were better techniques
to produce the analytic expression obtained by summing over all solutions. To overcome this difficulty, many methods have been proposed from various directions \cite{Kalousios:2015fya,Cardona:2015eba,Cardona:2015ouc,Dolan:2015iln,Huang:2015yka,Sogaard:2015dba,Bosma:2016ttj,
Zlotnikov:2016wtk,Cachazo:2015nwa,Gomez:2016bmv,Cardona:2016bpi,Baadsgaard:2015voa,Baadsgaard:2015ifa}.
Among these approaches, one of the most
efficient ways is the integration rules proposed by Baadsgaard, Bjerrum-Bohr, Bourjaily
and Damgaard \cite{Baadsgaard:2015voa,Baadsgaard:2015ifa}.
Inspired by the computation of amplitudes in the field theory limit of string theory,
they derived a simple set of combinatorial rules which immediately give the result after integration
for any M\"obius invariant integrand involving simple poles only. One can get the
desired final result after integration by applying these rules directly rather than solving
scattering equations. However, the requirement that the CHY-integrand contains only simple
poles could not be satisfied in general. Logically, there are two alternative issues to
bypass this disadvantage. One is to search the integration rules for higher-order poles \cite{Huang:2016zzb},
the other is to decompose an integrand of higher-order poles into
that of simple poles \cite{Baadsgaard:2015voa,Bjerrum-Bohr:2016juj}.

Recently, an algorithm of solving this problem has been
proposed by Cardona, Feng, Gomez and Huang, based on the
so-called cross-ratio identities \cite{Cardona:2016gon}. These identities reflect the relations between
rational functions in terms of $z_{ij}$ with different structures of poles.
By applying the cross-ratio identities iteratively, a systematic
algorithm for reducing CHY-integrands with higher-order poles has been established.
After decomposing the integrand into terms with simple poles only,
one can compute the desired analytic result via the integration rules.
This is the first systematic way to
get the analytic expression of an amplitude from a generic CHY-integrand.

Although this algorithm can be applied to any M\"obius invariant integrand
in principle, an important question is, can it be terminated
within finite steps for any CHY-integrand? In \cite{Bjerrum-Bohr:2016axv},
it has been proved that any weight-two rational function of $z_{ij}$
can be decomposed as a sum of PT-factors with kinematic coefficients
via the cross-ratio identities within finite steps.
Since any term from a known CHY-integrand can be expressed as a product of two
weight-two rational functions, one can conclude that all known
CHY-integrands can be decomposed into terms of only simple poles
by applying the cross-ratio identity method.

To verify its validity, it is worth appling this new method to integrands of
various theories and checking the result numerically. In this paper we consider the
following theories: non-linear sigma model (NLSM), special Galileon theory (SG), pure
Yang-Mills theory (YM), pure gravity (GR), Born-Infeld theory (BI),
Dirac-Born-Infeld theory (DBI) and its extension,
Yang-Mills-scalar theory (YMS), Einstein-Maxwell theory
(EM), Einstein-Yang-Mills theory (EYM).
All known CHY-integrands involving higher-order poles
are contained in the cases above.
In the meanwhile, theories corresponding to CHY-integrands
with simple poles only, such as
the scalar theory with $\phi^3$ or $\phi^4$ interaction,
will not be considered in this paper.
We divide them into three classes according to different building blocks of integrands.
The first class includes NLSM, SG, YM, BI as well as GR. Integrands of these theories
can be constructed from a $2n\times2n$ antisymmetric matrix $\Psi$.
The second class includes DBI, EM and a special case of YMS
of which integrands depend on antisymmetric
matrices $[\Psi]_{a, b : a}$, $[{\cal X}]_b$ as well as $\Psi$. The third class includes the general
YMS, the extended DBI and EYM, which contains the mixed traces of the generators of Lie groups.
Integrands of these theories are related to a polynomial ${\sum_{\{i,j\}}}'{\cal P}_{\{i,j\}}$,
or equivalently, an antisymmetric matrix $\Pi$. Computation shows
that all amplitudes considered in this paper can be calculated
efficiently within finite steps.

This paper is organized as follows: In section \eref{cross-ratio} we give a brief
review of the cross-ratio identity method. Based on this approach,
calculations of amplitudes of theories
in the three classes above are given in sections \eref{1class},
\eref{2class} and \eref{3class} respectively.
Finally, we give a brief summary in section \eref{conclusion}.

\section{Brief review of the cross-ratio identity method\label{cross-ratio}}

For reader's convenience, we will give
a brief introduction to the cross-ratio
identity method in this section \cite{Cardona:2016gon},
then we will discuss its validity
for general CHY-integrands.

\subsection{The systematic decomposition algorithm}

Firstly, we need to define
the order of poles of an integrand.
A generic $n$-point CHY-integrand consists of terms as rational
functions of $z_{ij}$ in the form
\bea
{\cal I}={1\over \prod _{i,j\in\{1,2,\ldots,n\},i<j}z_{ij}^{\beta_{ij}}}\,,
\eea
with integer power $\beta_{ij}$'s under the constraint
of the M\"obius invariance: $\sum_j\beta_{ij}+\sum_j\beta_{ji}=4$
for arbitrary $i\in\{1,2,\ldots,n\}$. For a
subset $\Lambda=\{i_1,i_2,\ldots,i_{|\Lambda|}\}\subset\{1,2,\ldots,n\}$,
the pole index $\chi_{\Lambda}$ is defined as
\bea
\chi_{\Lambda}=\Big(\sum_{i',j'\in \Lambda}\beta_{i'j'}\Big)-2(|\Lambda|-1)\,,
\eea
where $|\Lambda|$ denotes the length of the set $\Lambda$.
If $\chi_{\Lambda}\geq0$, a pole ${1\over s_{\Lambda}^{\chi+1}}$ will arise
in the result. It is straightforward to verify $\chi_{\Lambda}=\chi_{\bar{\Lambda}}$
which reflects the momentum conservation constraint $s_{\Lambda}=s_{\bar{\Lambda}}$,
where the subset $\bar{\Lambda}=\{1,2,\ldots,n\}\setminus\Lambda$ is the complement of $\Lambda$.
Thus, it is necessary to choose independent $\Lambda$'s.
If a CHY-integrand has $m$ independent subsets $\Lambda_1,\Lambda_2,\ldots,\Lambda_m$
with $\chi_{\Lambda_i}\geq0$, the order of poles of the integrand is
defined as
\bea
{\Upsilon}[{\cal I}]=\sum_{i=1}^m \chi_{\Lambda_i}\,.
\eea
Then an integrand which result in
simple poles only must satisfy $\Upsilon[{\cal I}]=0$.

In order to apply the integration rules,
it is necessary to decompose an integrand
with $\Upsilon[{\cal I}]>0$ into terms
with $\Upsilon[{\cal I'}]=0$. This
can be achieved by multiplying the cross-ratio
identities to the integrand iteratively.
The cross-ratio identity for the set $\Lambda$ is given by
\bea
\boxed{1=-\sum_{i\in\Lambda\setminus\{j\}}\sum_{b\in\bar{\Lambda}\setminus\{p\}}{s_{ib}\over s_\Lambda}
{z_{bp}z_{ij}\over z_{ib}z_{jp}}\equiv\mathbb{I}_n[\Lambda,j,p]\,,}
\eea
where $j$ and $p$ are selected manually.
This identity holds on the support of the scattering
equations and the momentum conservation constraint. One can expand the original ${\cal I}$
into $(|\Lambda|-1)(n-|\Lambda|-1)$ terms
via the operation ${\cal I}=\mathbb{I}_n[\Lambda,j,p]{\cal I}$.
This operation will not
break the manifest M\"obius invariance since
the cross-ratio identity is weight-zero under M\"obius transformations
for any node $i$ with $i\in\{1,2,\ldots,n\}$.
Obviously, such an operation decreases $\chi_\Lambda$ by $1$,
thus the order of the pole ${1\over s_\Lambda^{\chi+1}}$
has been reduced if $\chi_\Lambda>0$.

The systematic reduction algorithm is presented in the following:
\begin{enumerate}
  \item Count the order of poles $\Upsilon[\mathcal{I}]$ of the integrand ${\cal I}$. If
  $\Upsilon[\mathcal{I}]>0$, find the full set of independent subsets with
  $\chi_{\W\Lambda_i}>0$ (say, there are $m$ subsets):
  \bea
  \W\Lambda_1\,,~~\W\Lambda_2\,,~~\ldots\,,~~\W\Lambda_{m}\,.~~~\eea
  \item Step 1: start from the first set $\W\Lambda_1$, collect all $|\W\Lambda_1|(n-|\W\Lambda_1|)$ cross-ratio
  identities of $\W\Lambda_1$ with different choices of $j$ and $b$ as
  \bea \mathbb{I}_n[\W\Lambda_1,j,p]~~~~\mbox{where}~~~~j\in
  \W\Lambda_1\,,~~p\in
  \{1,2,\ldots,n\}\setminus\W\Lambda_1\,.~~~\eea
  \item Step 2: decompose the CHY-integrand ${\cal I}$ by applying the first
  cross-ratio identity
  \bea {\cal I}=\mathbb{I}_n[\W\Lambda_1,j,p]{\cal I}
  =\sum_{\ell}
  c_{\ell}{\cal I}'_{\ell}\,,\eea
  where
  $c_{\ell}$'s are rational functions of Mandelstam variables.
  \item Count all $\Upsilon[{\cal I}'_\ell]$.
  \begin{itemize}
  \item If all $\Upsilon[{\cal I}'_\ell]<\Upsilon[{\cal I}]$,
  return $\sum_{\ell} c_{\ell}\mathcal{I}'_{\ell}$.
  \item If there exists any $\Upsilon[\mathcal{I}']\geq\Upsilon[\mathcal{I}]$, test
  the second cross-ratio identity in step 2 and so on, until we
  find a cross-ratio identity satisfying all
  $\Upsilon[{\cal I}'_\ell]<\Upsilon[{\cal I}]$.
  \item If after
  running over all cross-ratio identities of the set
  $\Lambda_1$, there is still no such an identity satisfying all $\Upsilon[{\cal I}'_\ell]<\Upsilon[{\cal I}]$,
  then take the first identity in step 2
  again but now we stop till an identity satisfies all
  $\Upsilon[{\cal I}'_\ell]\leq \Upsilon[{\cal I}]$, and return $\sum_{\ell}
  c_{\ell}{\cal I}'_{\ell}$.
  \end{itemize}
\item Perform the procedure above for each ${\cal I}'_\ell$, and repeat the same
operation iteratively, then end with the expression such that the order of poles
of each term is zero.
\item If after some steps, there always exist terms with the order of poles
no less than $\Upsilon[{\cal I}]$, restart the algorithm by starting
from $\W\Lambda_2$, etc.

\end{enumerate}
The entire algorithm can be implemented in {\sc Mathematica}.
For a given CHY-integrand, if this algorithm can be terminated within finite steps,
one can obtain an expression with terms that have simple poles only, and finally get the
analytic result by applying the integration rules.

\subsection{The feasibility of this method}

Given the decomposition algorithm,
it is natural to ask whether all M\"obius
invariant integrands can be computed in this
way. It is obvious that the algorithm can
be performed for any integrand. The
question is, can it be terminated within finite steps?
An ideal situation is, we can always find
cross-ratio identities such that all terms satisfy $\Upsilon[{\cal I}'_\ell]<\Upsilon[{\cal I}]$
at each step, then the decomposition procedure
can be terminated after $\Upsilon[{\cal I}]$
steps at most. However, this happens for some
particular integrands rather than for all.
Thus, we need to consider if it is possible that at every step
there are terms carrying the structure of poles such
that $\Upsilon[{\cal I}'_\ell]=\Upsilon[{\cal I}]$,
or even $\Upsilon[{\cal I}'_\ell]>\Upsilon[{\cal I}]$,
for all choices of $\W\Lambda_i$, $j$ and $p$.
If this happens, the corresponding integrand cannot
be calculated by the method introduced
in the previous subsection. In order to fully understand the
cross-ratio identity method,
one needs to prove that the situation above
can be excluded in general, or
clarify when such a situation might arise.

Actually, the sum of $\chi_\Lambda$'s for
length-$t$ subsets of $\{1,2,\ldots,n\}$
is fully determined by the condition $\sum_{j\in\{1,2,\ldots,n\}}\beta_{ij}=4$ as
\bea
\chi_t\equiv\sum_{|\Lambda_i|=t}\chi_{\Lambda_i}=-2(t-1)C_n^2+2n\,C_{n-2}^{n-t}\,.
\eea
Thus, the sum of all $\chi_\Lambda$'s $\chi_{\rm total}\equiv\sum_{\Lambda} \chi_\Lambda=\sum_t \chi_t$ is invariant under
any action which maintains the M\"obius invariance. If
$\chi_{\rm total}$ is positive for some integer $n$,
it is impossible to decompose the corresponding integrand into terms
with simple poles only. Fortunately,
a little algebra leads to the conclusion that $\chi_t>0$ if
and only if $n<t+1$, thus $\chi_{\rm total}$ can never be positive.

On the other hand, it has been proved that a weight-two
rational function of $z_{ij}$ can always be transformed
to the sum of PT-factors ${1\over z_{i_1i_2}z_{i_2i_3}\cdots z_{i_ni_1}}$'s with kinematic coefficients
via the cross-ratio identities within finite steps \cite{Bjerrum-Bohr:2016axv}.
Any term of a known CHY-integrand in the literature can be
expressed as a product of two weight-two rational
functions. Hence, although it is not clear whether the
CHY-integrand for any physical theory can be decomposed as products of weight-two functions,
one can use the cross-ratio identities
to decompose any known CHY-integrand into terms
which contain simple poles only.

In this paper, we will choose the original algorithm rather
than the one which decomposes a weight-two function
into PT-factors, since the former is more convenient to be implemented
in {\sc Mathematica}. Indeed, the feasibility of this algorithm
has not been proved, since the procedure of decomposing
a weight-two function into PT-factors cannot ensure $\Upsilon[{\cal I}'_\ell]\leq\Upsilon[{\cal I}]$
at each step. However, as can be seen in the subsequent sections,
all known CHY-integrands can be computed by the original
algorithm efficiently, i.e., the condition $\Upsilon[{\cal I}'_\ell]\leq\Upsilon[{\cal I}]$
can always be satisfied, at least for all known CHY-integrands.

\section{Amplitudes of theories in the first class\label{1class}}

For theories in this class, the most important object in the construction of the
integrand ${\cal I}_n$ is the $2n\times 2n$ anti-symmetric matrix
\begin{equation}\label{Psi}
\Psi = \left(
         \begin{array}{c|c}
           ~~A~~ &  -C^{\rm T} \\
           \hline
           C & B \\
         \end{array}
       \right)
\end{equation}
where $A$, $B$ and $C$ are $n\times n$ matrices given by
\bea
A_{ij} = \begin{cases} \displaystyle {k_{i}\cdot k_j\over z_{ij}} & i\neq j\,,\\
\displaystyle  ~~~ 0 & i=j\,,\end{cases} \qquad B_{ij} = \begin{cases} \displaystyle {\epsilon_i\cdot\epsilon_j\over z_{ij}} & i\neq j\,,\\
\displaystyle ~~~ 0 & i=j\,,\end{cases}
\label{ABmatrix}
\eea
and
\bea
C_{ij} = \begin{cases} \displaystyle {\epsilon_i \cdot k_j\over z_{ij}} &\quad i\neq j\,,\\
\displaystyle -\sum_{l=1,\,l\neq i}^n\hspace{-.5em}{\epsilon_i \cdot k_l\over z_{il}} &\quad i=j\,.\end{cases}
\eea
One also needs to introduce the reduced Pfaffian ${\rm Pf}'\Psi={(-)^{i+j}\over z_{ij}}{\rm Pf}\Psi_{ij}^{ij}$
where $\Psi_{ij}^{ij}$ denotes the minor obtained by deleting rows and columns labeled by $i$ and $j$,
with $i,j\in\{1,2,\ldots , n\}$.
On the support of scattering equations, the reduced Pfaffian ${\rm Pf}'\Psi$
is invariant with respect to the permutation of particle labels.
In addition, a useful factor is defined as
\bea
{\cal C}_{\{i_1,i_2,\ldots,i_s\}}=\sum_{\sigma \in S_s/{\mathbb{Z}_s}}{{\rm Tr}(T^{I_{\sigma(i_1)}}T^{I_{\sigma(i_2)}}\cdots T^{I_{\sigma(i_s)}})\over z_{\sigma(i_1)\sigma(i_2)},z_{\sigma(i_2)\sigma(i_3)}\cdots z_{\sigma(i_s)\sigma(i_1)}}\,,
\eea
where $T^I$'s are generators of the Lie group under consideration.

The diagonal terms of the matrix $C$ will break the manifest
M\"obius invariance, since they are not of a uniform
weight under M\"obius transformations. To keep the validity
of the integration rules,
they need to be rewritten as
\bea
C_{ii}=\sum_{l\neq i}\Big({\epsilon_i\cdot k_l\over z_{ai}}+{(\epsilon_i\cdot k_l)z_{la}\over z_{ai}z_{il}}\Big)
\Rightarrow \sum_{l\neq i,a}{(\epsilon_i\cdot k_l)z_{la}\over z_{ai}z_{il}}\,,~~~~~~a\neq i\,,
\eea
where momentum conservation and the gauge invariant condition $\epsilon_i\cdot k_i=0$
have been used. The new formula of $C_{ii}$ gives the weight two for node $i$
and weight zero for other nodes, then the term-wise M\"obius invariance is
guaranteed. Throughout this paper, we choose
\bea
C_{ii} = \begin{cases} \displaystyle \sum_{l=3}^n {(\epsilon_1\cdot k_l)z_{l2}\over z_{21}z_{1l}}&\quad i=1\,,\\
\displaystyle \sum_{l\neq1,i}{(\epsilon_i\cdot k_l)z_{l1}\over z_{1i}z_{il}}&\quad i>1\,.\end{cases}\label{cd}
\eea
With these ingredients, we can now investigate theories in the first
class one by one.

\subsection{Non-linear sigma model}

We begin with the simplest case, the NLSM, whose standard Lagrangian
in Cayley parametrization is
\bea
{\cal L}^{{\tiny\mbox{NLSM}}}={1\over 8\la^2}{\rm Tr}(\partial_\mu {\rm U}^\dag\partial^\mu {\rm U})\,,
\eea
where
\bea
{\rm U}=(\mathbb{I}+\lambda \Phi)\,(\mathbb{I}-\lambda\Phi)^{-1}\,,~~~\Phi=\phi_I T^I\,.~~\label{U}
\eea
Here $\mathbb{I}$ is the identity matrix and $T^I$'s are generators of $U(N)$.
The CHY-integrand of NLSM is given by \cite{Cachazo:2014xea}
\bea
{\cal I}^{{\tiny\mbox{NLSM}}}=\lambda^{n{-}2}\,{\cal C}_n\,({\rm Pf}' A(k,z))^2\,.
\eea
For this case, it is sufficient to calculate the color-ordered partial
amplitude ${\cal A}^{{\rm NLSM}}(1,2,\ldots , n)$ defined via
\bea
{\cal A}^{{\tiny\mbox{NLSM}}}_n=\sum_{\sigma \in S_n/{\mathbb{Z}_n}}{\rm Tr}(T^{I_{\sigma(1)}}T^{I_{\sigma(2)}}\cdots T^{I_{\sigma(n)}})
{\cal A}^{{\tiny\mbox{NLSM}}}(\sigma(1),\sigma(2),\ldots ,\sigma(n))\,.
\eea
In other words, we focus on the color-ordered partial integrand
\bea
{\cal I}^{{\tiny\mbox{NLSM}}}(1,2,\ldots , n)={({\rm Pf}' A(k,z))^2\over z_{12}z_{23}\cdots z_{n1}}\,.
\eea
Here the coupling constant have been omitted.

We start from the 6-point amplitude ${\cal A}^{{\tiny\mbox{NLSM}}}_6$.
By definition, the corresponding color-ordered partial integrand is
\bea
{\cal I}^{{\tiny\mbox{NLSM}}}(1,2\ldots ,6)&=&{(k_1\cdot k_2)^2(k_3\cdot k_4)^2\over z_{12}^3z_{23}z_{34}^3z_{45}z_{56}^3z_{61}}
+{2(k_1\cdot k_2)(k_2\cdot k_4)(k_2\cdot k_3)(k_3\cdot k_4)\over z_{12}^2z_{14}z_{23}^2z_{34}^2z_{45}z_{56}^3z_{61}}\nn
& &-{2(k_1\cdot k_2)(k_1\cdot k_3)(k_2\cdot k_4)(k_3\cdot k_4)\over z_{12}^2z_{13}z_{23}z_{24}z_{34}^2z_{45}z_{56}^3z_{61}}
+{(k_1\cdot k_4)^2(k_2\cdot k_3)^2\over z_{12}z_{14}^2z_{23}^3z_{34}z_{45}z_{56}^3z_{61}}\nn
& &+{(k_1\cdot k_3)^2(k_2\cdot k_4)^2\over z_{12}z_{13}^2z_{23}z_{24}^2z_{34}z_{45}z_{56}^3z_{61}}
-{2(k_1\cdot k_3)(k_1\cdot k_4)(k_2\cdot k_3)(k_2\cdot k_4)\over z_{12}z_{13}z_{14}z_{23}^2z_{24}z_{34}z_{45}z_{56}^3z_{61}}\,.
~~~~\label{I-NLSM6}
\eea
The pole structure of \eref{I-NLSM6} is listed as follows:
\begin{center}
\begin{tabular}{|c|c|c|c|c|c|c|c|c|}
  \hline
  ~ & 1st~term & 2nd~term & 3rd term & 4th~term & 5th~term & 6th~term \\
  \hline
  pole & ${1\over s_{12}^2s_{34}^2s_{56}^2s_{123}s_{612}s_{561}}$ & ${1\over s_{56}^2s_{12}s_{23}s_{34}s_{123}s_{561}}$
  & ${1\over s_{56}^2s_{12}s_{34}s_{123}s_{561}}$ &${1\over s_{23}^2s_{56}^2s_{14}s_{123}s_{561}}$
  & ${1\over s_{56}^2s_{13}s_{24}s_{123}s_{561}}$ & ${1\over s_{56}^2s_{23}s_{123}s_{561}}$ \\
  \hline
\end{tabular}
\end{center}
It can be seen from the table that every term contains higher-order poles
which need to be decomposed. Via the cross-ratio identity method,
One can accomplish the decomposition within three steps.
Below is the table with $\#[\mbox{ALL}]$, the
number of resulting terms and $\#[\mbox{H}]$, the number
of terms of higher-order poles in each {\sl Round} of decomposition:
\begin{center}
\begin{tabular}{|c|c|c|c|c|c|c|c|c|}
  \hline
  ~ & Round 1 & Round 2 & Round 3  \\
  \hline
  $\#[\mbox{ALL}]$ & 18 & 30 & 38 \\
  \hline
  $\#[\mbox{H}]$ & 6 & 4 & 0 \\
  \hline
\end{tabular}
\end{center}
Integrations of these terms can be bypassed by applying the integration rules.
Summing all terms from the final result, we obtain
\bea
{\cal A}^{{\tiny\mbox{NLSM}}}(1,2,\ldots ,6)&=&{(s_{12}+s_{23})(s_{45}+s_{56})\over s_{123}}+{(s_{23}+s_{34})(s_{56}+s_{61})\over s_{234}}
+{(s_{34}+s_{45})(s_{61}+s_{12})\over s_{345}}\nn
& &-(s_{12}+s_{23}+s_{34}+s_{45}+s_{56}+s_{61}).\label{nlsm1}
\eea
For this simple example, the full computation takes less than a minute in {\sc Mathematica}.
One can see the manifest cyclic symmetry in \eref{nlsm1},
which is
the characteristic of the color-ordered partial
amplitude.
This analytic result
is confirmed numerically against
the one obtained from solving scattering
equations numerically.

Then we turn to the 8-point amplitude
${\cal A}^{{\tiny\mbox{NLSM}}}_8$. The integrand has
$120$ terms and all terms contain higher-order
poles. Performing the cross-ratio identity method,
this integrand can be decomposed into
$4340$ terms with simple poles only within $6$ steps.
The table of $\#[\mbox{ALL}]$
and $\#[\mbox{H}]$ in each {\sl Round} of decomposition
is given as follows:
\begin{center}
\begin{tabular}{|c|c|c|c|c|c|c|c|c|}
  \hline
  ~ & Round 1 & Round 2 & Round 3 & Round 4 & Round 5 & Round 6 \\
  \hline
  $\#[\mbox{ALL}]$ & 600 & 1128 & 1904  & 2924 & 4084 & 4340 \\
  \hline
  $\#[\mbox{H}]$ & 124 & 142 & 169 & 150 & 32 & 0 \\
  \hline
\end{tabular}
\end{center}
From these terms with simple poles, we get the desired analytic
expression of the amplitude via the integration rules.
The final result can be simplified into the form
\bea
{\cal A}^{{\tiny\mbox{NLSM}}}(1,2,\ldots ,8)={\rm Part}1-{\rm Part}2+{\rm Part}3\,,
\eea
with
\bea
{\rm Part}1&=&{(s_{12}+s_{23})(s_{45}+s_{56})(s_{78}+s_{8123})\over s_{123}s_{456}}
+{(s_{23}+s_{34})(s_{56}+s_{67})(s_{81}+s_{1234})\over s_{234}s_{567}}\nn
& &+{(s_{34}+s_{45})(s_{67}+s_{78})(s_{12}+s_{2345})\over s_{345}s_{678}}
+{(s_{45} + s_{56}) (s_{78}+ s_{81}) (s_{23} + s_{3456})\over s_{456} s_{781}}\nn
& &+{(s_{56} + s_{67})(s_{81} + s_{12})  (s_{34} + s_{4567})\over s_{567} s_{812}}
+{(s_{67}+s_{78})(s_{12}+s_{23})(s_{45}+s_{5678})\over s_{678}s_{123}}\nn
& &+{(s_{78}+s_{81})(s_{23}+s_{34})(s_{56}+s_{6789})\over s_{781}s_{234}}
+{(s_{81}+s_{12})(s_{34}+s_{45})(s_{67}+s_{7812})\over s_{812}s_{345}}\nn
& &+{(s_{12} + s_{23}) (s_{56} + s_{67}) (s_{1234} +s_{4567})\over s_{123} s_{567}}
+{(s_{23} + s_{34}) (s_{67} + s_{78}) (s_{2345} +s_{5678})\over s_{234} s_{678}}\nn
& &+{(s_{34} + s_{45}) (s_{78} + s_{81}) (s_{3456} +s_{6781})\over s_{345} s_{781}}
+{(s_{45} + s_{56}) (s_{81} + s_{12}) (s_{4567} +s_{7812})\over s_{456} s_{812}}\,,
\eea
\bea
{\rm Part}2&=&{(s_{12} + s_{23}) (s_{45} + s_{56} + s_{67} + s_{78}+ s_{8123} +s_{1234} )\over s_{123}}\nn
& &+{(s_{23} + s_{34}) (s_{56} + s_{67} + s_{78} + s_{81} + s_{1234}+s_{2345} )\over s_{234}}\nn
& &+{(s_{34} + s_{45}) (s_{67} + s_{78} + s_{81} + s_{12} + s_{2345}+s_{3456} )\over s_{345}}\nn
& &+{(s_{45} + s_{56}) (s_{78} + s_{81} + s_{12} + s_{23} + s_{3456}+s_{4567} )\over s_{456}}\nn
& &+{(s_{56} + s_{67}) (s_{81} + s_{12} + s_{23} + s_{34} + s_{4567}+s_{5678} )\over s_{567}}\nn
& &+{(s_{67} + s_{78}) (s_{12} + s_{23} + s_{34} + s_{45} + s_{5678}+s_{6781} )\over s_{678}}\nn
& &+{(s_{78} + s_{81}) (s_{23} + s_{34} + s_{45} + s_{56} + s_{6781}+s_{7812} )\over s_{781}}\nn
& &+{(s_{81} + s_{12}) (s_{34} + s_{45} + s_{56} + s_{67} + s_{7812}+s_{8123} )\over s_{812}}\,,
\eea
\bea
{\rm Part}3&=&2(s_{12}+s_{23}+s_{34}+s_{45}+s_{56}+s_{67}+s_{78}+s_{81})+s_{1234}+s_{2345}+s_{3456}+s_{4567}\,.
\eea
This result has been verified numerically.

\subsection{Special Galileon theory}

The next theory under consideration is SG.
The general pure Galileon Lagrangian is
\bea
{\cal L}^{\tiny\mbox{SG}}=-{1\over2}\,\partial_\mu\phi\,\partial^\mu\phi+\sum_{m=3}^\infty g_m\,{\cal L}_m\,,
\eea
with
\bea
{\cal L}_m=\phi\,{\rm det}\,\{\partial^{\mu_i}\,\partial_{\nu_j}\phi\}_{i,j=1}^{m-1}\,.
\eea
We restrict ourselves on the special situation in which there exist
constraints on coupling constants $g_m$'s
such that all amplitudes with an odd number
of external particles vanish. Then the CHY-integrand
${\cal I}_n$ of this theory is given by \cite{Cachazo:2014xea}
\bea
{\cal I}^{\tiny\mbox{SG}}=({\rm Pf}' A(k,z))^4\,,
\eea
where the coupling constants have been omitted.

With this setup, we choose the 6-point
amplitude ${\cal A}^{\tiny\mbox{SG}}_6$ as an example. The integrand
has $15$ terms, and all terms contain
higher-order poles. One can divide it into
$3169$ terms with simple poles only within $10$ steps.
The table of $\#[\mbox{ALL}]$
and $\#[\mbox{H}]$ in each {\sl Round} of decomposition
is given by
\begin{center}
\begin{tabular}{|c|c|c|c|c|c|c|c|c|}
  \hline
  ~ & Round 1 & Round 2 & Round 3 & Round 4 & Round 5  \\
  \hline
  $\#[\mbox{ALL}]$ & 45 & 108 & 234  & 468 & 873  \\
  \hline
  $\#[\mbox{H}]$ & 45 & 72 & 162 & 180 & 198   \\
  \hline
\end{tabular}
\end{center}
\begin{center}
\begin{tabular}{|c|c|c|c|c|c|c|c|c|}
  \hline
  ~ & Round 6 & Round 7 & Round 8 & Round 9 & Round 10  \\
  \hline
  $\#[\mbox{ALL}]$ & 1215 & 1944 & 3068  & 3151 & 3169  \\
  \hline
  $\#[\mbox{H}]$ & 261 & 530 & 53 & 12 & 0   \\
  \hline
\end{tabular}
\end{center}
Although the final result is too lengthy to be presented, it has been confirmed numerically.

\subsection{Yang-Mills theory}

Then we turn to the pure YM. The CHY-integrand
of YM is \cite{Cachazo:2014xea}
\bea
{\cal I}^{\tiny\mbox{YM}}={\cal C}_n\,{\rm Pf}'\Psi(k,\epsilon,z)\,.
\eea
Similar to the case of NLSM, it is sufficient
to consider the color-ordered partial integrand
\bea
{\cal I}^{\tiny\mbox{YM}}(1,2,\ldots , n)={{\rm Pf}'\Psi(k,\epsilon,z)\over z_{12}z_{23}\cdot\cdot\cdot z_{n1}}\,.
\eea

Let us take the 6-point color-ordered amplitude
${\cal A}^{\tiny\mbox{YM}}(1,2,\ldots ,6)$ as an example.
The partial integrand ${\cal I}^{\tiny\mbox{YM}}(1,2,\ldots , 6)$
has $3420$ terms and $1120$ of them
contain higher-order poles.
The decomposition procedure can be terminated
within $5$ steps via the cross-ratio
identity method, and the analytic
expression of ${\cal A}^{\tiny\mbox{YM}}(1,2,\ldots ,6)$
is verified numerically.
The table of $\#[\mbox{ALL}]$
and $\#[\mbox{H}]$ in each {\sl Round} of decomposition
is given by
\begin{center}
\begin{tabular}{|c|c|c|c|c|c|c|c|c|}
  \hline
  ~ & Round 1 & Round 2 & Round 3 & Round 4 & Round 5  \\
  \hline
  $\#[\mbox{ALL}]$ & 6174 & 8624 & 10459 & 11167 & 11252  \\
  \hline
  $\#[\mbox{H}]$ & 834 & 533 & 162 & 28 & 0   \\
  \hline
\end{tabular}
\end{center}

It is worth noticing that, when checking
the result numerically, the values of external momenta
must satisfy the momentum conservation
constraint, which is necessary for the derivation of the cross-ratio identities.
However, those of polarization vectors
can be chosen arbitrarily since they are irrelevant to
the cross-ratio identities
and the integration. We have verified
the result with polarization vectors $\epsilon_i\cdot k_i\neq0$
as well as $\epsilon_i\cdot k_i=0$,
and find that the analytic expression reproduces the value
obtained from solving scattering equations numerically.

\subsection{Born-Infeld theory}

Now we consider BI whose Lagrangian is given by
\bea
{\cal L}^{\tiny\mbox{BI}}=\ell^{-2}\,\Big(\sqrt{-{\rm det}(\eta_{\mu\nu}-\ell^2F_{\mu\nu})}-1\Big)\,.
\eea
The CHY-integrand of BI is \cite{Cachazo:2014xea}
\bea
{\cal I}^{\tiny\mbox{BI}}= \ell^{n{-}2}\,{\rm Pf}' \Psi(k,\tilde\epsilon, z)\,{\rm Pf}' A(k,z)^2\,.
\eea

For simplicity, we calculate the 6-point
amplitude ${\cal A}^{\tiny\mbox{BI}}_6$. The integrand
contains $20400$ terms and $18744$ of them
involve higher-order poles. Using the cross-ratio
identities, one can reduce it to terms
with simple poles within $10$ steps. The table of $\#[\mbox{ALL}]$
and $\#[\mbox{H}]$ in each {\sl Round} of decomposition
is given by
\begin{center}
\begin{tabular}{|c|c|c|c|c|c|c|c|c|}
  \hline
  ~ & Round 1 & Round 2 & Round 3& Round 4& Round 5 \\
  \hline
  $\#[\mbox{ALL}]$ & 61200 & 123267 & 202067 & 269132 &324740    \\
  \hline
  $\#[\mbox{H}]$ & 28616 & 32813 &24206 &15026 &6420   \\
  \hline
\end{tabular}
\end{center}
\begin{center}
\begin{tabular}{|c|c|c|c|c|c|c|c|c|}
  \hline
  ~ & Round 1 & Round 2 & Round 3& Round 4& Round 5 \\
  \hline
  $\#[\mbox{ALL}]$ & 352236 & 375220 & 397206 & 399183 & 399552    \\
  \hline
  $\#[\mbox{H}]$ & 3173 & 3135 & 304 & 62 & 0  \\
  \hline
\end{tabular}
\end{center}
This is the most complicated example in this paper, which takes more than a day
in {\sc Mathematica}. The analytic expression of the amplitude
is confirmed by numerical verification.

\subsection{Gravity}

The final theory under consideration in this section is
GR. The CHY-integrand of this theory is the
product of two independent copies of the one for YM,
each of which has its own gauge choice for polarization
vectors \cite{Cachazo:2014xea}
\bea
{\cal I}^{\tiny\mbox{GR}}={\rm Pf}'\Psi(k,\epsilon,z)\,{\rm Pf}'\Psi(k,\tilde{\epsilon},z)\,.
\eea
The polarization tensor of a graviton
is given by $\zeta_{\mu\nu}=\epsilon_\mu\tilde{\epsilon}_\nu$.
This integrand leads to amplitudes of gravitons coupled to dilatons
and B-fields.

We take the 4-point amplitude
${\cal A}^{\tiny\mbox{GR}}_4$ as an example. The integrand contains $484$ terms,
with $228$ terms involving higher-order poles. It can be decomposed
into terms with simple poles within $2$ steps, as shown in
the following table
\begin{center}
\begin{tabular}{|c|c|c|c|c|c|c|c|c|}
  \hline
  ~ & Round 1 & Round 2  \\
  \hline
  $\#[\mbox{ALL}]$ & 484 & 484   \\
  \hline
  $\#[\mbox{H}]$ & 36 & 0   \\
  \hline
\end{tabular}
\end{center}
Physically, polarization tensors
of gravitons are traceless, i.e., they satisfy
$\epsilon^\mu\tilde{{\epsilon}}_\mu=0$. However,
as discussed before, their values can be chosen
without imposing any physical constraint when performing
the numerical verification.

\section{Amplitudes of theories in the second class\label{2class}}

In this section we move on to theories in the second class.
CHY-integrands of these theories require
two new matrices $[{\cal X}]_b$
and $[\Psi]_{a, b : a}$ as basic ingredients.
$a$ and $b$ are two sets of external particles,
whose numbers are denoted by $n_a$ and $n_b$ respectively,
and $n=n_a+n_b$ is the total number of particles.
$[{\cal X}]_b$ is an $n_b\times n_b$ matrix
defined as
\bea
([{\cal X}]_b)_{i,j}=\begin{cases} \displaystyle \frac{\delta^{I_i, I_j}}{z_{ij}} & i\neq j\,,\\
\displaystyle ~ ~ 0 & i=j\,,\end{cases}
\eea
where $I^i\in\{1,...,M\}$ denotes the $i$-th $U(1)$ charge of the $U(1)^M$ group.
$[\Psi]_{a, b : a}$ is an $(n+n_a)\times (n+n_a)$
matrix obtained from $\Psi$ by deleting rows and
columns labeled by $n+i$ for all $i\in [n_a+1,n]$.
More explicitly, its elements are given by
\bea
([\Psi]_{a, b : a})_{i,j}=\begin{cases} \displaystyle ~~A_{ij} & i,j\in \{1,2,\ldots , n\}\,,\\
\displaystyle ~~C_{(i-n)j} & i\in \{n+1,n+2,\ldots , n+n_a\},~j\in \{1,2,\ldots , n\}\,,\\
\displaystyle ~~(-C^T)_{i(j-n)} & i\in \{1,2,\ldots , n\},~j\in \{n+1,n+2,\ldots , n+n_a\}\,,\\
\displaystyle ~~B_{(i-n)(j-n)} &i,j\in \{n+1,n+2,\ldots , n+n_a\}\,.  \end{cases}
\eea
Among $n$ external particles, only
particles in subset $a$ contribute
their polarization vectors to the matrix $[\Psi]_{a, b : a}$.
This is the reason why the integrand including ${\rm Pf}' [\Psi]_{a, b : a}$
can describe interactions between bosons with different spins.

It is worth emphasizing that all terms in ${\rm Pf}[{\cal X}]_b$
are manifestly invariant under the M\"obius transformations. From
the formula of diagonal terms of the matrix $C$ in \eref{cd}, terms in
the expansion of ${\rm Pf}'[\Psi]_{a, b : a}$
also have the manifest M\"obius invariance,
which ensures the feasibility of the integration rules.

\subsection{Special Yang-Mills-Scalar theory}

The first theory under consideration in this section
is a special case of YMS which describes the low energy
effective action of $N$ coincident D-branes.
The Lagrangian of this theory is
\bea
{\cal L}^{\tiny\mbox{YMS}}=-{\rm Tr} \Big({1\over4} F^{\mu \nu} F_{\mu \nu} + {1\over2} D^\mu \phi^I\,D_\mu \phi^I- {{g^2}\over 4} \sum_{I\neq J} [\phi^I, \phi^J]^2 \Big)\,,
\eea
where the gauge group is $U(N)$, and the scalars carry a flavor index $I$
with $I\in\{1,...,M\}$ from the $M$-dimensional
space transverse to the D-brane.
The corresponding CHY-integrand is \cite{Cachazo:2014xea}
\bea
{\cal I}^{\tiny\mbox{YMS}}(g,s)={\cal C}_n\,{\rm Pf} [{\cal X}]_s(z)\,{\rm Pf}' [\Psi]_{g, s: g}(k,\tilde\epsilon, z)\,,
\eea
where $g$ and $s$ denote the sets of gluons and
scalars respectively.
Gluons have polarization vectors $\epsilon^\mu$'s
while scalars do not, thus their kinematical information can be combined
into the matrix $[\Psi]_{g, s: g}$.
Again, we consider the color-ordered
partial amplitude ${\cal A}^{\tiny\mbox{YMS}}(1,2,\ldots , n)$.

The first example is the 6-point partial amplitude
${\cal A}^{\tiny\mbox{YMS}}(1g,2g,3g,4g,5s,6s)$, where external particles
$1g$, $2g$, $3g$ and $4g$ are gluons, while
$5s$, $6s$ are scalars of the same flavor.
The partial integrand has $222$ terms and $68$
of them contain higher-order poles.
The decomposition procedure is shown
in the following table
\begin{center}
\begin{tabular}{|c|c|c|c|c|c|c|c|c|}
  \hline
  ~ & Round 1 & Round 2 & Round 3   \\
  \hline
  $\#[\mbox{ALL}]$ & 376 & 514 & 592   \\
  \hline
  $\#[\mbox{H}]$ & 51 & 22 & 0   \\
  \hline
\end{tabular}
\end{center}

The second example is the 6-point amplitude
${\cal A}^{\tiny\mbox{YMS}}(1g,2g,3s_{I_1},4s_{I_1},5s_{I_2},6s_{I_2})$,
where $1g$ and $2g$ are gluons,
$3s_{I_1}$, $4s_{I_1}$ are scalars
of one flavor and $5s_{I_2}$, $6s_{I_2}$
are scalars of another.
The partial integrand contains
$15$ terms and $7$ of them contain
higher-order poles. The decomposition
can be done within $3$ steps as shown in the following table
\begin{center}
\begin{tabular}{|c|c|c|c|c|c|c|c|c|}
  \hline
  ~ & Round 1 & Round 2 & Round 3   \\
  \hline
  $\#[\mbox{ALL}]$ & 29 & 35 & 43   \\
  \hline
  $\#[\mbox{H}]$ & 3 & 4 & 0   \\
  \hline
\end{tabular}
\end{center}

Analytic expressions of these two examples
are verified numerically.

\subsection{Dirac-Born-Infeld theory}

We proceed to consider DBI whose Lagrangian is
\bea
{\cal L}^{\tiny\mbox{DBI}}=\ell^{-2}\,\Big(\sqrt{-{\rm det}(\eta_{\mu\nu}-\ell^2\,\partial_\mu\phi^I\,\partial_\nu\phi^I-\ell F_{\mu\nu})}-1\Big)\,,
\eea
where $I$ again labels the flavor of scalars.
The CHY-integrand of DBI is \cite{Cachazo:2014xea}
\bea
{\cal I}^{\tiny\mbox{DBI}}(\gamma,s)=\ell^{n{-}2}\,{\rm Pf} [{\cal X}]_s(z)\,{\rm Pf}' [\Psi]_{\gamma,s: \gamma}(k,\tilde\epsilon, z)\,({\rm Pf}' A(k,z))^2\,,
\eea
where $\gamma$ denotes the set of photons and $s$ the set of scalars respectively.

Let us calculate
the 6-point amplitude ${\cal A}^{\tiny\mbox{DBI}}_{2\gamma2s_{I_1}2s_{I_2}}$
which contains two photons and
four scalars carrying two flavor indices.
The integrand has $82$
terms and all of them contain
higher-order poles. One can accomplish the
decomposition procedure via the cross-ratio
identities within $10$ steps, as shown
in the following table
\begin{center}
\begin{tabular}{|c|c|c|c|c|c|c|c|c|}
  \hline
  ~ & Round 1 & Round 2 & Round 3 & Round 4 & Round 5  \\
  \hline
  $\#[\mbox{ALL}]$ & 246 & 440 & 674  & 904 & 1169 \\
  \hline
  $\#[\mbox{H}]$ & 106 & 120 & 130 & 104 & 67   \\
  \hline
\end{tabular}
\end{center}
\begin{center}
\begin{tabular}{|c|c|c|c|c|c|c|c|c|}
  \hline
  ~ & Round 6 & Round 7 & Round 8 & Round 9 & Round 10  \\
  \hline
  $\#[\mbox{ALL}]$ & 1285 & 1528 & 1908  & 1937 & 1943  \\
  \hline
  $\#[\mbox{H}]$ & 87 & 174 & 17 & 4 & 0   \\
  \hline
\end{tabular}
\end{center}
Again, this result is verified numerically.

\subsection{Einstein-Maxwell theory}

The final theory in this section is EM
which describes gravitons coupled to
photons. The CHY-integrand of this theory
is given by \cite{Cachazo:2014xea}
\bea
{\cal I}^{\tiny\mbox{EM}}={\rm Pf} [{\cal X}]_\gamma(z)\,{\rm Pf}' [\Psi]_{h, \gamma : h}(k,\epsilon,  z)\,{\rm Pf}' \Psi(k,\tilde\epsilon, z)\,.~~\label{integrand-EM}
\eea
where the set of gravitons is denoted by $h$,
and that of photons is denoted by $\gamma$.
The expression \eref{integrand-EM} allows the photons to carry more than one flavor in general.
The polarization tensor of a graviton is $\zeta^{\mu\nu}
=\epsilon^\mu\tilde\epsilon^\nu$, and the polarization vector of a photon is $\tilde\epsilon^\nu$.
The matrix  $\Psi(k,\tilde\epsilon, z)$ contains $\tilde\epsilon^\nu$ for both gravitons and photons,
and the matrix $[\Psi]_{h, \gamma : h}(k,\epsilon,  z)$ contains the remaining $\epsilon^\mu$'s
for gravitons.

Our example is the 5-point amplitude ${\cal A}^{\tiny\mbox{EM}}_{3h2\gamma}$
whose external particles are three gravitons and two photons carrying the same flavor index.
The integrand has $5013$ terms and $1171$ of
them contain higher-order poles.
The decomposition procedure can be done
within $4$ steps, as shown in the following table
\begin{center}
\begin{tabular}{|c|c|c|c|c|c|c|c|c|}
  \hline
  ~ & Round 1 & Round 2 & Round 3 & Round 4   \\
  \hline
  $\#[\mbox{ALL}]$ & 6748 & 7577 & 7775  & 7799  \\
  \hline
  $\#[\mbox{H}]$ & 488 & 80 & 10 & 0   \\
  \hline
\end{tabular}
\end{center}
Again, we have verified this result numerically.

\section{Amplitudes of theories in the third class\label{3class}}

Theories in the final
class correspond to multi-trace
mixed amplitudes. More precisely, an
amplitude in this section contains
external bosons which belong to the set $a\cup b_{{\rm Tr}_1}\cup b_{{\rm Tr}_2}\cdots \cup b_{{\rm Tr}_m}$
with bosons in set $a$ of spin $S_{a}$
and that in set $b_{{\rm Tr}_i}$'s of
spin $S_b=S_a-1$.
This structure leads to the mixed color factor
${{\rm Tr}_1}\cup {\rm Tr}_2\cdots \cup {\rm Tr}_m$
in the amplitude.
The kinematic part
of the integrand for these theories can be constructed
through two equivalent ways, one is to introduce the polynomial
${\sum_{\{i,j\}}}'{\cal P}_{\{i,j\}}$,
the other is to define the matrix $\Pi$.
Let us assume the amplitude contains
$m$ mixed traces, then ${\sum_{\{i,j\}}}'{\cal P}_{\{i,j\}}$
is a sum over the perfect
matching $\{i,j\}$
\bea \label{piexpansion}
{\sum_{\{i,j\}}}'{\cal P}_{\{i,j\}}=\sum_{i_1<j_1\in{\rm Tr}_1}^{i_{m-1}<j_{m-1}\in{\rm Tr}_{m-1}}{\rm sgn}(\{i,j\})\,
z_{i_1 j_1}\cdots z_{i_{m-1} j_{m-1}}\,
{\rm Pf}[\Psi]_{a,i_1,j_1,\ldots,i_{m-1},j_{m-1}: h}\,,
\eea
where $i_a$ and $j_a$ are labels of two external particles
which belong to $b_{{\rm Tr}_a}$.
This sum can be recognized as the
reduced Pfaffian of the matrix $\Pi$.
The matrix $\Pi$ can be constructed
from $\Psi$ by performing the so-called squeezing
operation iteratively.
Terms in the expansion of ${\sum_{\{i,j\}}}'{\cal P}_{\{i,j\}}$
respect the M\"obius invariance automatically, while
terms in the expansion of ${\rm Pf}'\Pi$ break
the manifest M\"obius invariance thus are forbidden
for the integration rules. Hence, we will
use ${\sum_{\{i,j\}}}'{\cal P}_{\{i,j\}}$
to express integrands throughout
this section.

\subsection{General Yang-Mills-Scalar theory}

Let us consider the general YMS with the Lagrangian
\bea
{\cal L}^{\tiny\mbox{gen.YMS}}=-{\rm Tr} \Big({1\over4} F^{\mu \nu} F_{\mu \nu} + {1\over2} D^\mu \phi^I\, D_\mu \phi^I- {{g^2}\over 4} \sum_{I\neq J} [\phi^I, \phi^J]^2 \Big)+{\lambda\over 3!}f_{IJK}f_{\bar{I}\bar{J}\bar{K}}\phi^{I\bar{I}}\phi^{J\bar{J}}\phi^{K\bar{K}}\,,
\eea
which involves the general flavor group and a cubic scalar self-interaction.
The trace is for the gauge group, and $f_{\bar{I}\bar{J}\bar{K}}$ and $f_{IJK}$ are the structure constants of gauge and flavor groups respectively. Amplitudes of this theory can only contain a single trace of the gauge group, and multi-traces for the flavor group,
as can be seen from the general CHY-integrand \cite{Cachazo:2014xea}
\bea
{\cal I}^{\tiny\mbox{gen.YMS}}(s_{{\rm Tr}_1}\cup\cdots\cup s_{{\rm Tr}_m}, g)={\cal C}_n\,{\cal C}_{{\rm Tr}_1}\cdots {\cal C}_{{\rm Tr}_m}
{\sum_{\{i,j\}}}'{\cal P}_{\{i,j\}}(s_{{\rm Tr}_1}\cup\cdots\cup s_{{\rm Tr}_m}, g)\,,
\eea
where $g$ denotes the set of gluons and $s_{{\rm Tr}_i}$ denotes
the set of scalars with the trace ${\rm Tr}_i$.

Obviously, the simplest example is that the scalars belong to two traces and
each trace contains two scalars. However, these amplitudes correspond
to the special case of YMS with ${\rm Tr}(T^{I_{i_1}}T^{I_{i_2}})$ replaced
by $\delta^{I_{i_1},I_{i_2}}$.
In other words, the kinematic part of these amplitudes is included
in the situations we have calculated in the previous section.
Thus, we choose to compute a non-trivial case,
the 7-point color-ordered partial amplitude
${\cal A}^{\tiny\mbox{gen.YMS}}(1g,(2s,3s,4s)_{{\rm Tr}_1},(5s,6s,7s)_{{\rm Tr}_2})$,
which contains one gluon and six scalars
with three scalars carrying ${\rm Tr}_1$ and the rest three
carrying ${\rm Tr}_2$.
The integrand has $21$ terms and all of them contain higher-order poles.
The decomposition procedure can be done within $2$ steps, as
shown in the following table
\begin{center}
\begin{tabular}{|c|c|c|c|c|c|c|c|c|}
  \hline
  ~ & Round 1 & Round 2  \\
  \hline
  $\#[\mbox{ALL}]$ & 126 & 216   \\
  \hline
  $\#[\mbox{H}]$ & 18 & 0   \\
  \hline
\end{tabular}
\end{center}
Again, this analytic result is confirmed by
the numerical verification.

\subsection{Extended Dirac-Born-Infeld theory}

The second theory under consideration is the extended
DBI, which is described by the Lagrangian
\bea
{\cal L}^{\tiny\mbox{ext.DBI}}=\ell^{-2}\,\Big(\sqrt{-{\rm det}\big(\eta_{\mu\nu}-{\ell^2\over 4\la^2}{\rm Tr}
(\partial_\mu {\rm U}^\dag\,\partial_\nu {\rm U})-\ell^2\,W_{\mu\nu}-\ell F_{\mu\nu}\big)}-1\Big)\,,
\eea
where the matrix ${\rm U}(\Phi)$ is defined in \eref{U}, and
\bea
W_{\mu\nu}=\sum_{m=1}^{\infty}\,\sum_{k=0}^{m-1}
\,{2(m-k)\over 2m+1}\,\la^{2m+1}\,{\rm Tr}(\partial_{[\mu}\Phi\,\Phi^{2k}\,\partial_{\nu]}\Phi\,\Phi^{2(m-k)-1})\,.
\eea
The corresponding CHY-integrand is given by \cite{Cachazo:2014xea}
\bea
{\cal I}^{\tiny\mbox{ext.DBI}}(s_{{\rm Tr}_1}\cup\cdots\cup s_{{\rm Tr}_m}, \gamma)={\cal C}_{{\rm Tr}_1}\cdots {\cal C}_{{\rm Tr}_m}
{\sum_{\{i,j\}}}'{\cal P}_{\{i,j\}}(s_{{\rm Tr}_1}\cup\cdots\cup s_{{\rm Tr}_m}, \gamma)\,({\rm Pf}'A)^2\,,
\eea
where $\gamma$ denotes the set of photons and
$s_{{\rm Tr}_i}$ denotes the set of scalars
with the trace ${\rm Tr}_i$.

Our example is the 6-point partial amplitude
${\cal A}^{\tiny\mbox{ext.DBI}}(1\gamma,(2s,3s)_{{\rm Tr}_1}(4s,5s,6s)_{{\rm Tr}_2})$,
which involves one photon and five scalars,
where two scalars carry ${\rm Tr}_1$
and three carry ${\rm Tr}_2$.
The integrand has $36$ terms
and $8$ of them contain higher-order
poles. The decomposition procedure
can be done within $3$ steps as shown
in the table
\begin{center}
\begin{tabular}{|c|c|c|c|c|c|c|c|c|}
  \hline
  ~ & Round 1 & Round 2  & Round 3 \\
  \hline
  $\#[\mbox{ALL}]$ & 52 & 88 & 142  \\
  \hline
  $\#[\mbox{H}]$ & 18 & 18 & 0 \\
  \hline
\end{tabular}
\end{center}
This result has been verified numerically.

\subsection{Einstein-Yang-Mills theory}

The final theory in this section is the
Einstein-Yang-Mills theory, which describes
the interaction between gravitons and
gauge bosons. The general CHY-integrand
involving the mixed traces is \cite{Cachazo:2014xea}
\bea
{\cal I}^{\tiny\mbox{EYM}}(g_{{\rm Tr}_1}\cup\cdots\cup g_{{\rm Tr}_m},h)={\cal C}_{{\rm Tr}_1}\cdots {\cal C}_{{\rm Tr}_m}
{\sum_{\{i,j\}}}'{\cal P}_{\{i,j\}}(g_{{\rm Tr}_1}\cup\cdots\cup g_{{\rm Tr}_m}, h)\,{\rm Pf}'\Psi\,,
\eea
where the set of gravitons is denoted by $h$ and the set
of gluons with the trace ${\rm Tr}_i$ is denoted by $g_{{\rm Tr}_i}$.

We consider the 5-point partial amplitude ${\cal A}^{\tiny\mbox{EYM}}((1g,2g)_{{\rm Tr}_1}(3g,4g,5g)_{{\rm Tr}_2})$,
of which all five external particles are gluons
with two of them carrying ${\rm Tr}_1$ and three
of them carrying ${\rm Tr}_2$. The original integrand
has $239$ terms and $189$ of them contain
higher-order poles. The decomposition procedure
can be done within $3$ steps, as shown in the
following table
\begin{center}
\begin{tabular}{|c|c|c|c|c|c|c|c|c|}
  \hline
  ~ & Round 1 & Round 2  & Round 3 \\
  \hline
  $\#[\mbox{ALL}]$ & 434 & 531 & 557  \\
  \hline
  $\#[\mbox{H}]$ & 115 & 26 & 0 \\
  \hline
\end{tabular}
\end{center}
As all other examples, this analytic
result is confirmed numerically.

\section{Conclusion\label{conclusion}}

In this paper, we have applied the cross-ratio
identity method to CHY-integrands of various theories
including: non-linear sigma model, special Galileon
theory, pure Yang-Mills theory, pure gravity,
Born-Infeld theory, Dirac-Born-Infeld theory and
its extension,
Yang-Mills-scalar theory, Einstein-Maxwell theory
as well as Einstein-Yang-Mills theory.
All the integrands under consideration are computed
conveniently in this way, the decomposition
procedures expend $10$ steps at most. All the analytic
results are verified numerically, thus this method
is confirmed for all examples of this paper.
Consequently, the cross-ratio identity method
is valid and effective for a wide range of CHY-integrands.
An interesting observation is that the condition
$\Upsilon[{\cal I}'_\ell]\leq\Upsilon[{\cal I}]$
can always be satisfied at each step, although its
rigorous proof is still absent.

In this paper, the most complicated example takes more than a day in {\sc Mathematica}.
The reason of this low efficiency is, the choices of $\W\Lambda_i$, $j$ and $p$ are
tested by brute force in the algorithm.
Appropriate choices of the cross-ratio identities at
each step can minimize the number of steps of the decomposition, which
is crucial for practical calculations.
Thus, how to optimize
these choices to improve the efficiency
is an important future project.

{\bf Acknowledgment }

The author would like to thank Prof. Bo Feng and Rijun Huang for their
help throughout the project. It is also necessary to thank Junjie Rao
for his carefully reading of the original manuscript. Finally, the author
would also like to acknowledge the supporting from Chinese Postdoctoral Administrative
Committee.

{}

\end{document}